\documentclass[preprint,aps]{revtex4}
\usepackage{amsmath}
\usepackage{amssymb}

\input{tcilatex}

\begin{document}

\title{The hierarchical Green function approach to the two-dimensional
Hubbard model}
\author{Yu-Liang Liu}
\affiliation{Department of Physics, Renmin University, Beijing 100872, \\
People's Republic of China}

\begin{abstract}
By introducing multipe-site correlation functions, we propose a hierarchical
Green function approach, and apply it to study the characteristic properties
of a 2D square lattice Hubbard model by solving the equation of motions of a
one-particle Green function and related multipe-site correlation functions.
Under a cut-off approximation and taking the Fourier representation of
multipe-site correlation functions, we obtain an analytical expression of
one-particle Green function with static correlation functions. Then we
calculate the spectral density function of electrons, and obtain that
besides two main peaks corresponding to the lower and upper Hubbard bands in
the spectral density function, there emerge some novel states between these
two main peaks, and the total spectral weight of these emerged states is
proportional to the hole doping concentration $\delta $. Meanwhile, there
also emerge some collective modes related to possible charge/spin density
wave and/or electronic pairing density wave ordering states. This
calculation is completely consistent with the spectroscopy observations of
the cuprate superconductors in normal states. On the other hand, the
appearence of the static correlation functions in the one-particle Green
function can be used to describe the intertwined orders observed in the
normal state of the cuprate superconductors.

74.72.-h, 03.65.Db, 71.27.+a.
\end{abstract}

\maketitle

\section{\protect\bigskip Introduction}

Since the discovery of the high $T_{c}$ cuprate superconducting materials%
\cite{1}, it is gradually realized that the strong correlation effect of
electrons play a key role in understanding of the normal and superconducting
states of these materials\cite{2,3,4,5}. Up to now there are a lot of
experimental data and numerical simulations showing that the novel behavior
of the normal states in the underdoped and optimal doped regimes of these
materials\cite{5a,6,7} originates from the strong correlation of electrons
produced by the strong repulsive Coulomb interaction of electrons, and these
unprecedented properties cannot be unambiguously explained by usual
perturbation theory of quantum many particle systems based on the
"independent particle" (quasi-particle) assumption of the Landau Fermi
liquid theory\cite{8}.

According to the present variety of experimental observations, the following
aspacts have been confirmed that: (1) the "parent" of the high $T_{c}$
cuprate superconductors is an antiferromagnetic Mott insulator (with a 1.5eV
charge-transfer energy gap) where the C$_{u}$-O plane(s) of the cuprate
materials is undoped, and there is one electron on each C$_{u}$ site; Hole
doping rapidly suppresses the antiferromagnetic N\'{e}el order, but the
optical gap does not collapse. The infrared reflectivity studies demonstrate
the coexistence of the charge-transfer gap with finite optical conductivity
that is transferred into the gap. Integrating the conductivity within the
gap, the effective carrier density grows in proportion to the hole doping
concentration\cite{8ad1,8ad2,8ad3} $\delta $, rather than $1-\delta $ as
predicted by conventional band theory. (2) As the hole doping concentration $%
\delta $ reaching around $\delta _{\min }\simeq 5\%$, superconductivity sets
in, with a transition temperature $T_{c}$ that grows to a maximum at about $%
\delta _{opt}\simeq 16\%$, optimal doping, then declines for higher dopings
and vanishes for $\delta _{\max }\simeq 27\%$. The superconducting state has
a dome in the phase diagram of temperature versus hole doping level for
cuprates. Materials with $\delta _{\min }<\delta <\delta _{opt}$ and $\delta
_{opt}<\delta <\delta _{\max }$ are referred to, respectively, as
"underdoped" and "overdoped". For the enough large hole doping $\delta
_{\max }\leq \delta $, the superconductivity disappears, and the system can
be represented by usual Landau Fermi liquid theory where the quasiparticle
concept is established. Moreover, the superfluid density scales
approximately with the hole doping concentration $\delta $ in the underdoped
regime\cite{88ad1}. Obviously, the parameter of the hole doping
concentration $\delta $ play a key role in understanding of low temperature
physical behavior of the cuprates, and it really sets up a bridge between an
antiferromagnetic Mott insulator and the Landau Fermi liquid. (3) In the
underdoped regime, at temperature just above $T_{c}$, the normal state has a
"pseudogap" which is characterized by a substantial suppression of the
electronic density of states at low energies that cannot be simply related
to the occurence of any form of broken symmetry, within which there are
strong and ubiquitous tendencies toward several sorts of order with similar
energy scales, including various forms of charge density wave (CDW), spin
density wave (SDW), and possibly pairing density wave (PDW). These
"intertwined orders" make the normal state of the cuprates show ineluctable
complexity\cite{8ad4}, and remain to be understood.

The Hubbard model and the related t-J model are widely thought to capture
the essential physics of a class of highly correlated systems, such as the
high T$_{c}$ cuprate superconductors. The two-dimensional (2D) Hubbard model%
\cite{7a} in a square lattice, used to describe the basic characters of high 
$T_{c}$ cuprate supercunductivity\cite{5a,6,7,8a,88a}, has been extensively
studied in both analytical and numerical calculations, where there is
inherent frustration between the tendency to maintain local
antiferromagnetic correlations originated from the on-site strong repulsive
Coulomb interaction and the doped hole itineracy.

The effective treatment of the influence of the on-site repulsive Coulomb
interaction on the states of electrons is a central issue of any theoretical
approach, where at the large repulsive $U$, a double occupied state on each
site is strongly suppressed, and the Hilbert space of the electrons is split
into two subspaces: one is composed of the unoccupied and single occupied
states, and another one composed of the double occupied states that are
lifted up high energy levels. In fact, there emerges a single-occupied
constraint condition for electrons on each site produced by the on-site
strong repulsive Coulomb interaction, which is a major difficulty faced by
the present approaches. On the other hand, it is well known that in the both
cases of weak $U/t_{0}\ll 1$ and strong $U/t_{0}\rightarrow \infty $
coupling limits, where $t_{0}$ is the hopping amplitude of electrons, the
basic property of the ground state of the 2D square lattice Hubbard model is
clear: in the former it is a Fermi liquid\cite{81} as apart from the half
filling; and in the latter it is a fully polarized ferromagnetic metallic
phase\cite{82} away from the half filling, in which there does not appear
any order state.

The rich physical phenomena shown by the 2D square lattice Hubbard model
really appear in the intermediate coupling, where $U$ is of order the
bandwidth $W$($=8t_{0}$), $U\sim W$, where there is the keen competition
between the kinetic energy and the on-site repulsive Coulomb interaction of
electrons. The former takes the delocalization of electrons, while the
latter makes electrons localize. In this coupling range, there is still not
a ubiquitous acceptable calculations from microscopic theories. The 2D
square lattice Hubbard model with intermediate coupling, likely cannot be
treated using any fundamentally perturbative approach which starts with a
non-interacting particle description. Beyond the present perturbation
theoretical methods, the on-site Coulomb interaction of electrons had to be
treated effectively before taking any approximation in analytical and
numerical calculations.

In this paper, by introducing multipe-site correlation functions, we
originally conceive a hierarchic Green function approach (HGFA) to study the
equation of motion (EOM) of a one-particle Green function, in which the
contribution of the electron correlation effect produced by the Coulomb
interaction is completely represented by high order multipe-site correlation
functions. In this way, by including the multipe-site correlation functions,
the influence of the on-site repulsive Coulomb interaction on the states of
electrons can be effectively incorporated into the one-particle Green
function.

The idea of the HGFA is that, in contrast with the EOM of Green function
approach that is usually written out in the phase space\cite{10,11,12,13,13a}%
, we write out the EOMs of the one-particle Green function and related high
order multipe-site correlation functions in the lattice space for the 2D
square lattice Hubbard model, then we take the possible cut-off
approximations for the highest order multipe-site correlation functions
emerging in these EOMs. The two salient features of the HGFA are that: one
is that the contribution coming from the on-site repulsive Coulomb
interaction term can be rigorously incorporated into the EOMs, which is
really represented by related hierarchical multipe-site correlation
functions, and another one is that these hierarchical multipe-site
correlation functions appearing in these EOMs can be classified into
different levels denoted by a parameter $L$ which labels the number of
electrons residing in a length (sites) scale that take part in a time
evolution of an electron from initial state to final state. The\
multipe-site correlation functions in the same level $L$ constitute a set of
linear EOMs, and in these EOMs only a few related multipe-site correlation
functions belonging to the level $L+1$ are appearing. In the lattice
representation, the physical picture of a multipe-site correlation function
in the level $L$ is very clear, that it represents the time evolution of an
electron from the site $\mathbf{x}_{i}$ at time $t_{1}$ to the site $\mathbf{%
x}_{j}$ at time $t_{2}$ with considering the influence by the number $L$ of
other electrons around this site $\mathbf{x}_{i}$ at time $t_{1}$. It is
expectant that for the higher $L$, the role of the multipe-site correlation
functions in the level $L$ is weaker. This character of the multipe-site
correlation functions is very helpful in taking the cut-off approximations
for the highest order correlation functions appearing in the EOMs. After the
cut-off approximation, this set of EOMs of the one-particle Green function
and the related multipe-site correlation functions is completely closed, and
it can be solved after taking the Fourier representation of these
multipe-site correlation functions.

Applying the HGFA for the 2D square lattice Hubbard model with an
intermediate coupling $U$, we reveal that the one-particle Green function of
electrons has the following salient characters: (1) The excitation energy
spectrum of electrons is split into two subbands for a large $U$, called the
lower and upper Hubbard bands, and the gap between these two bands is linear
proportion to $U$; At half filling the system is a Mott insulator; As the
hole doping, there emerge new states within this gap, and their total
spectral weight is proportional to the hole doping concentration $\delta $;
However, as taking enough large hole doping, $\delta _{\max }\leq \delta $,
the emerged states can fill in this gap. (2) The series of hierarchical
static correlation functions are emergent in the one-particle Green
function, which originate from the Coulomb interaction of electrons, and can
be used to represent the contributions from possible intertwined orders
appearing in the underdoped regime.

For the 2D square lattice Hubbard model, the HGFA in fact describes the
intrinsic competition between the kinetic and potential energies of
electrons which induces the inherent frustration between the tendency to
maintain local antiferromagnetic correlations and the doped hole itineracy.
As for both cases of $U=0$ and $W=0$, the HGFA is rigorous. In other case,
the on-site Coulomb interaction can produce new multipe-site correlation
functions belonging to the level $L+1$ in the EOM of a multipe-site
correlation function in the level $L$ by increasing a local factor $U%
\widehat{n}_{i\sigma }$ in a corresponding site $\mathbf{x}_{i}$, where $%
\widehat{n}_{i\sigma }$ is the number operator of electrons, while the
kinetic energy part of electrons is responsible to make each correlation
function have a "hopping term", and to connect each EOM of a related
multipe-site correlation function with others in the same level $L$ to
construct a set of linear EOMs. The physics picture of the HGFA is very
clear that the on-site Coulomb interaction plays the role to establish the
relation between the multipe-site correlation functions in the level $L$ and
those related ones belonging to the level $L+1$, while the kinetic energy
part only connects the multipe-site correlation functions in the same level $%
L$. Consequently there naturally appears a series of the hierarchical linear
EOMs of related multipe-site correlation functions. Moreover, in this series
of the hierarchical linear EOMs there emerge a series of hierarchical static
correlation functions that are related to the corresponding multipe-site
correlation functions. These static correlation functions can be used to
describe the inhomogeneous states appearing in the system, and they can
drastically influence the spectral density function of electrons.

The organization of the paper is as follows: we give the detail description
of the HGFA in section II by using the EOM of an one-particle Green function
for the Hubbard model by writing out a few EOMs of the related multipe-site
correlation functions with $L=2$. In section III, after taking the Fourier
representation of the related multipe-site correlation functions, we solve
this set of EOMs under the cutt-off approximations for the multipe-site
correlation functions belonging to the level $L=3$, and give an analytical
expression of a one-particle Green function. In section IV, we calculate the
electronic spectral density function by the one-particle Green function, and
demonstrate that the total spectral weight of the emerged states within the
gap between the lower and upper Hubbard bands is proportional to a hole
doping concentration $\delta $. The conclusion will be given in Section V.

\section{The EOMs of the one-particle Green function and multipe-site
correlation functions}

The Hamiltonian of the 2D square lattice Hubbard model is that,%
\begin{equation}
\widehat{H}=-t_{0}\sum_{ij\sigma }\widehat{\gamma }_{ij}\left( \widehat{c}%
_{i\sigma }^{\dagger }\widehat{c}_{j\sigma }+\widehat{c}_{j\sigma }^{\dagger
}\widehat{c}_{i\sigma }\right) +U\sum_{i}\widehat{n}_{i\uparrow }\widehat{n}%
_{i\downarrow }-\mu \sum_{i\sigma }\widehat{n}_{i\sigma }  \label{1}
\end{equation}%
where $\widehat{c}_{i\sigma }^{\dagger }$ ($\widehat{c}_{i\sigma }$) creates
(annihilates) an electron with spin $\sigma =\uparrow $, $\downarrow $ on
site $\mathbf{x}_{i}$, $\widehat{n}_{i\sigma }=\widehat{c}_{i\sigma
}^{\dagger }\widehat{c}_{i\sigma }$ is the number operator, $\mu $ is the
chemical potential, $t_{0}$ is the hopping amplitude, and $U$ is the on-site
repulsive Coulomb interaction strength. The hopping factor $\widehat{\gamma }%
_{ij}$ is defined as that,%
\begin{equation*}
\widehat{\gamma }_{ij}=\left\{ 
\begin{array}{cc}
1, & j=i+1 \\ 
0, & j\neq i+1%
\end{array}%
\right.
\end{equation*}%
which denotes the summation over the sites $\mathbf{x}_{i},\mathbf{x}_{j}$
only in the nearest neighbor. In order to write the EOMs of a one-particle
Green function and related multipe-site correlation functions, we need the
following commutation relations, 
\begin{eqnarray}
\lbrack \widehat{c}_{i\sigma },\widehat{H}] &=&\sum_{j}\left( \widehat{h}%
_{ij}-\mu \delta _{ij}\right) \widehat{c}_{j\sigma }+U\widehat{n}_{i%
\overline{\sigma }}\widehat{c}_{i\sigma }  \notag \\
\lbrack \widehat{n}_{i\sigma },\widehat{H}] &=&\sum_{j}\widehat{h}_{ij}%
\widehat{X}_{ij\sigma }^{(-)}  \label{3} \\
\lbrack \widehat{X}_{ij\overline{\sigma }}^{(\pm )},\widehat{H}] &=&U\left( 
\widehat{n}_{j\sigma }-\widehat{n}_{i\sigma }\right) \widehat{X}_{ij%
\overline{\sigma }}^{(\mp )}-\sum_{l}\widehat{h}_{il}\widehat{X}_{lj%
\overline{\sigma }}^{(\mp )}+\sum_{l}\widehat{h}_{jl}\widehat{X}_{il%
\overline{\sigma }}^{(\mp )}  \notag
\end{eqnarray}%
where $\widehat{h}_{il}=-t_{0}\left( \widehat{\gamma }_{il}+\widehat{\gamma }%
_{li}\right) $, and the bonding operators $\widehat{X}_{ij\sigma }^{(\pm )}$
are defined as that, $\widehat{X}_{ij\sigma }^{(\pm )}=\widehat{c}_{i\sigma
}^{\dagger }\widehat{c}_{j\sigma }\pm \widehat{c}_{j\sigma }^{\dagger }%
\widehat{c}_{i\sigma }$. These commutation relations are the basic
ingredients of writing the EOMs of related multipe-site correlation
functions that directly or indirectly enter the EOM of the one-particle
Green function.

The one-particle Green function is defined as that,%
\begin{eqnarray}
G_{ij\sigma }(t_{1},t_{2}) &=&-i<\mathit{T}\widehat{c}_{i\sigma }(t_{1})%
\widehat{c}_{j\sigma }^{\dagger }(t_{2})>  \notag \\
&=&-i\theta (t_{1}-t_{2})<\widehat{c}_{i\sigma }(t_{1})\widehat{c}_{j\sigma
}^{\dagger }(t_{2})>  \label{6} \\
&&%
\begin{array}{ccc}
&  & +i\theta (t_{2}-t_{1})<\widehat{c}_{j\sigma }^{\dagger }(t_{2})\widehat{%
c}_{i\sigma }(t_{1})>%
\end{array}
\notag
\end{eqnarray}%
In the Heisenberg picture, the time dependent of the annihilation (creation)
operator $\widehat{c}_{i\sigma }(t)$($\widehat{c}_{i\sigma }^{\dagger }(t)$)
of an electron is represented by the Hamiltonian $\widehat{H}$ of the system
as that,%
\begin{eqnarray*}
\widehat{c}_{i\sigma }(t) &=&e^{i\widehat{H}t}\widehat{c}_{i\sigma }e^{-i%
\widehat{H}t} \\
\widehat{c}_{i\sigma }^{\dagger }(t) &=&e^{i\widehat{H}t}\widehat{c}%
_{i\sigma }^{\dagger }e^{-i\widehat{H}t}
\end{eqnarray*}%
As applied for the Hubbard model, the EOM of the one-particle Green function
reads that,%
\begin{equation}
\left[ \left( i\partial _{t_{1}}+\mu \right) \delta _{il}-\widehat{h}_{il}%
\right] G_{lj\sigma }(t_{1},t_{2})-UF_{iij\sigma }^{(n_{\overline{\sigma }%
})}(t_{1},t_{2})=\delta (t_{1}-t_{2})\delta _{ij}  \label{7}
\end{equation}%
where, $F_{iij\sigma }^{(n_{\overline{\sigma }})}(t_{1},t_{2})=-i<\mathit{T}%
\widehat{n}_{i\overline{\sigma }}(t_{1})\widehat{c}_{i\sigma }(t_{1})%
\widehat{c}_{j\sigma }^{\dagger }(t_{2})>$, is produced by the on-site
Coulomb interaction, and it is a multipe-site correlation function due to
the appearence of the number operator $\widehat{n}_{i\overline{\sigma }%
}(t_{1})$ in the corresponding site $\mathbf{x}_{i}$. Obviously, all
correlation effects of electrons in one-particle Green function is
completely coming from the contribution of the related correlation function $%
F_{iij\sigma }^{(n_{\overline{\sigma }})}(t_{1},t_{2})$. Physically, the
correlation function $F_{iij\sigma }^{(n_{\overline{\sigma }})}(t_{1},t_{2})$
in fact describes an evolution process of an electron from a double occupied
state on site $\mathbf{x}_{i}$ at time $t_{1}$ to a state on site $\mathbf{x}%
_{j}$ at time $t_{2}$.

In order to defining multipe-site correlation functions, we introduce new
composite multipe-site operators $\widehat{F}_{\{\alpha _{1}...\alpha
_{L}\}}^{(A_{1}\cdots A_{L})}$ that are composed of the density operators $%
\widehat{n}_{i\sigma }$ and the bonding operators $\widehat{X}_{il\sigma
}^{(\pm )}$,

\begin{equation}
\widehat{F}_{\{\alpha _{1}...\alpha _{L}\}}^{(A_{1}\cdots A_{L})}=\underset{%
k=1}{\Pi }^{L}\widehat{A}_{\alpha _{k}}  \label{5}
\end{equation}%
where $\widehat{A}_{\alpha }=\left\{ \widehat{n}_{i\overline{\sigma }},%
\widehat{n}_{i\sigma },\widehat{X}_{il\overline{\sigma }}^{(-)},\widehat{X}%
_{il\overline{\sigma }}^{(+)},\widehat{X}_{il\sigma }^{(-)},\widehat{X}%
_{il\sigma }^{(+)}\right\} $, and $L=N_{n_{\overline{\sigma }}}+N_{n_{\sigma
}}+N_{X_{\sigma }^{-}}+N_{X_{\sigma }^{+}}+N_{X_{\overline{\sigma }%
}^{-}}+N_{X_{\sigma }^{+}}$, where $N_{A}$ is the number of the operator $%
\widehat{A}_{\alpha }$ appearing in the composite operators $\widehat{F}%
_{\{\alpha _{1}...\alpha _{L}\}}^{(A_{1}\cdots A_{L})}$. With the help of
these composite operators $\widehat{F}_{\{\alpha _{1}...\alpha
_{L}\}}^{(A_{1}\cdots A_{L})}$, we define the corresponding multipe-site
correlation functions $F_{\{\alpha _{1}...\alpha _{L}\}mp\sigma
}^{(A_{1}\cdots A_{L})}(t_{1},t_{2})$, 
\begin{equation}
F_{\{\alpha _{1}...\alpha _{L}\}mp\sigma }^{(A_{1}\cdots
A_{L})}(t_{1},t_{2})=-i<\mathit{T}\widehat{F}_{\{\alpha _{1}...\alpha
_{L}\}}^{(A_{1}\cdots A_{L})}(t_{1})\widehat{c}_{m\sigma }(t_{1})\widehat{c}%
_{p\sigma }^{\dagger }(t_{2})>  \label{8}
\end{equation}%
Some of these multipe-site correlation functions $F_{\{\alpha _{1}...\alpha
_{L}\}mp\sigma }^{(A_{1}\cdots A_{L})}(t_{1},t_{2})$ will enter into the
series of hierarchical EOMs originated from Eq.(\ref{7}), and they will
construct a set of linear EOMs with the one-particle Green function $%
G_{ip\sigma }(t_{1},t_{2})$. The physics meaning of the operator $\widehat{F}%
_{\{\alpha _{1}...\alpha _{L}\}}^{(\widehat{A}_{1}\cdots \widehat{A}_{L})}%
\widehat{c}_{m\sigma }$ in the correlation function $F_{\{\alpha
_{1}...\alpha _{L}\}mp\sigma }^{(A_{1}\cdots A_{L})}(t_{1},t_{2})$ is that
an electron $\widehat{c}_{m\sigma }$ with spin $\sigma $ at site $\mathbf{x}%
_{m}$ attached other electron distribution represented by the operator $%
\widehat{F}_{\{\alpha _{1}...\alpha _{L}\}}^{(\widehat{A}_{1}\cdots \widehat{%
A}_{L})}$ around the site $\mathbf{x}_{m}$, where the parameter $L$ denotes
the number of electrons residing in a length (sites) scale around this site $%
\mathbf{x}_{m}$ that the electrons in this scale all are involved in the
time evolution process of the electron $\widehat{c}_{m\sigma }$. thus the
correlation function $F_{\{\alpha _{1}...\alpha _{L}\}mp\sigma
}^{(A_{1}\cdots A_{L})}(t_{1},t_{2})$ in fact represents the evolution
process of an electron from the initial state incorporated the influence of
a definite distribution of other electrons around this electron to final
state. As taking $L=1$ (i.e. $N_{n_{\overline{\sigma }}}=1$), we have the
correlation function $F_{ilj\sigma }^{(n_{\overline{\sigma }%
})}(t_{1},t_{2})=-i<\mathit{T}\widehat{n}_{i\overline{\sigma }}(t_{1})%
\widehat{c}_{l\sigma }(t_{1})\widehat{c}_{j\sigma }^{\dagger }(t_{2})>$,
that enters into the EOM of the one-particle Green function as taking $i=l$.

In contrast with usual correlation functions defined in the momentum space,
the present multipe-site correlation functions can more effectively describe
the correlation effect of electrons derived from the Coulomb interaction,
and the parameter $L$ appearing in the composite operators $\widehat{F}%
_{\{\alpha _{1}...\alpha _{L}\}}^{(\widehat{A}_{1}\cdots \widehat{A}_{L})}$
can be used to classify the correlation functions $F_{\{\alpha _{1}...\alpha
_{L}\}mp\sigma }^{(A_{1}\cdots A_{L})}(t_{1},t_{2})$ into different levels,
where the correlation functions $F_{\{\alpha _{1}...\alpha _{L}\}mp\sigma
}^{(A_{1}\cdots A_{L})}(t_{1},t_{2})$ in the same level $L$ can constitute
one or more subset of EOMs, in which only a few correlation functions
belonging to the level $L+1$ are emerging. For a large $L$ (i.e. a lot of
electrons in a large length scale take part in the time evolution of the
electron $\widehat{c}_{m\sigma }$), the contribution of the correlation
functions $F_{\{\alpha _{1}...\alpha _{L}\}mp\sigma }^{(A_{1}\cdots
A_{L})}(t_{1},t_{2})$ to the one-particle Green function is expectantly
small, therefor we can take the cut-off approximations for those belonging
to the level $L+1$ that appear in the EOMs of the correlation functions in
the level $L$, and we have a closed set of the EOMs of the one-particle
Green function and the related multipe-site correlation functions $%
F_{\{\alpha _{1}...\alpha _{L}\}mp\sigma }^{(A_{1}\cdots
A_{L})}(t_{1},t_{2}) $.

On the other hand, for a system with a few electron number $N_{e}$, the
number of the correlation functions $F_{\{\alpha _{1}...\alpha
_{L}\}mp\sigma }^{(A_{1}\cdots A_{L})}(t_{1},t_{2})$ is finite, because the
parameter $L$ in the multipe-site correlation functions $F_{\{\alpha
_{1}...\alpha _{L}\}mp\sigma }^{(A_{1}\cdots A_{L})}(t_{1},t_{2})$ must
satisfy the following condition,%
\begin{equation}
L\leq N_{e}  \label{12}
\end{equation}%
Therefore, the set of EOMs of the one-particle Green function $G_{lj\sigma
}(t_{1},t_{2})$ and the related multipe-site correlation functions $%
F_{\{\alpha _{1}...\alpha _{L}\}mp\sigma }^{(A_{1}\cdots
A_{L})}(t_{1},t_{2}) $ are automatically closed, and it can be
analytically/numerically solved.

According to the above definition of the multipe-site correlation functions
in Eq.(\ref{8}) and with the help of Eq.(\ref{3}), twe can write out he EOMs
of the correlation functions $F_{\{\alpha _{1}...\alpha _{L}\}mp\sigma
}^{(A_{1}\cdots A_{L})}(t_{1},t_{2})$ with $L=1$ that they enter the series
of the EOM of the one-particle Green function,%
\begin{eqnarray}
i\partial _{t_{1}}F_{ijq\sigma }^{(n_{\overline{\sigma }})}(t_{1},t_{2})
&=&\delta (t_{1}-t_{2})<\widehat{n}_{i\overline{\sigma }}>\delta
_{jq}+UF_{ijjq\sigma }^{(n_{\overline{\sigma }}n_{\overline{\sigma }%
})}(t_{1},t_{2})  \notag \\
&&+\sum_{m}\left( \widehat{h}_{jm}-\mu \delta _{jm}\right) F_{imq\sigma
}^{(n_{\overline{\sigma }})}(t_{1},t_{2})+\sum_{m}\widehat{h}%
_{im}F_{imjq\sigma }^{(X_{\overline{\sigma }}^{-})}(t_{1},t_{2})  \label{12a}
\end{eqnarray}%
\begin{eqnarray}
i\partial _{t_{1}}F_{iljq\sigma }^{(X_{\overline{\sigma }}^{\pm
})}(t_{1},t_{2}) &=&\delta (t_{1}-t_{2})<\widehat{X}_{il\overline{\sigma }%
}^{(\pm )}>\delta _{jq}+UF_{iljjq\sigma }^{(X_{\overline{\sigma }}^{\mp }n_{%
\overline{\sigma }})}(t_{1},t_{2})  \notag \\
&&+\sum_{m}\left( \widehat{h}_{jm}-\mu \delta _{jm}\right) F_{ilmq\sigma
}^{(X_{\overline{\sigma }}^{\pm })}(t_{1},t_{2})  \notag \\
&&-\sum_{m}\left[ \widehat{h}_{im}F_{mljq\sigma }^{(X_{\overline{\sigma }%
}^{\mp })}(t_{1},t_{2})-\widehat{h}_{lm}F_{imjq\sigma }^{(X_{\overline{%
\sigma }}^{\mp })}(t_{1},t_{2})\right]  \label{13} \\
&&+U\left( F_{liljq\sigma }^{(n_{\sigma }X_{\overline{\sigma }}^{\mp
})}(t_{1},t_{2})-F_{iiljq\sigma }^{(n_{\sigma }X_{\overline{\sigma }}^{\mp
})}(t_{1},t_{2})\right)  \notag
\end{eqnarray}%
where there emerge the new multipe-site correlation functions with $L=2$, $%
F_{niljq\sigma }^{(n_{\sigma }X_{\overline{\sigma }}^{\mp })}(t_{1},t_{2})$, 
$F_{ijjq\sigma }^{(n_{\overline{\sigma }}n_{\overline{\sigma }%
})}(t_{1},t_{2})$ and $F_{niljq\sigma }^{(X_{\overline{\sigma }}^{\pm }n_{%
\overline{\sigma }})}(t_{1},t_{2})$. In the $L=1$ level, the correlation
functions $F_{ijq\sigma }^{(n_{\overline{\sigma }})}(t_{1},t_{2})$ and $%
F_{iljq\sigma }^{(X_{\overline{\sigma }}^{\pm })}(t_{1},t_{2})$ constitute a
subset of EOMs, and they connect with the one-particle Green function $%
G_{iq\sigma }(t_{1},t_{2})$ by the correlation function $F_{iiq\sigma }^{(n_{%
\overline{\sigma }})}(t_{1},t_{2})$. On the other hand, other correlation
functions belonging to the $L=1$ level can be discarded because they do not
enter this series of the hierarchical EOMs of the one-particle Green
function and related multipe-site correlation functions.

To the next level $L=2$, we need to include all correlation functions that
enter this series of the hierarchical EOMs, such as, $F_{niljq\sigma
}^{(n_{\sigma }X_{\overline{\sigma }}^{\mp })}(t_{1},t_{2})$, $F_{ijjq\sigma
}^{(n_{\overline{\sigma }}n_{\overline{\sigma }})}(t_{1},t_{2})$ and $%
F_{niljq\sigma }^{(X_{\overline{\sigma }}^{\pm }n_{\overline{\sigma }%
})}(t_{1},t_{2})$, as well as that ones emerging in their EOMs. For example,
the EOM of the multipe-site correlation functions $F_{liljq\sigma
}^{(n_{\sigma }X_{\overline{\sigma }}^{\mp })}(t_{1},t_{2})$ can be written
as that,%
\begin{eqnarray}
i\partial _{t_{1}}F_{niljq\sigma }^{(n_{\sigma }X_{\overline{\sigma }}^{\pm
})}(t_{1},t_{2}) &=&\delta (t_{1}-t_{2})<\{\widehat{n}_{n\sigma }\widehat{c}%
_{j\sigma },\widehat{c}_{q\sigma }^{\dagger }\}\widehat{X}_{il\overline{%
\sigma }}^{\pm }>+UF_{niljjq\sigma }^{(n_{\sigma }X_{\overline{\sigma }%
}^{\pm }n_{\overline{\sigma }})}(t_{1},t_{2})  \notag \\
&&+\sum_{m}\left( \widehat{h}_{jm}-\mu \delta _{jm}\right) F_{nilmq\sigma
}^{(n_{\sigma }X_{\overline{\sigma }}^{\pm })}(t_{1},t_{2})  \notag \\
&&-\sum_{m}\left[ \widehat{h}_{im}F_{nmljq\sigma }^{(n_{\sigma }X_{\overline{%
\sigma }}^{\mp })}(t_{1},t_{2})-\widehat{h}_{lm}F_{nimjq\sigma }^{(n_{\sigma
}X_{\overline{\sigma }}^{\mp })}(t_{1},t_{2})\right]  \label{13a} \\
&&+\sum_{m}\widehat{h}_{nm}F_{nmiljq\sigma }^{(X_{\sigma }^{-}X_{\overline{%
\sigma }}^{\pm })}(t_{1},t_{2})  \notag \\
&&+U\left[ F_{nliljq\sigma }^{(n_{\sigma }n_{\sigma }X_{\overline{\sigma }%
}^{\mp })}(t_{1},t_{2})-F_{niiljq\sigma }^{(n_{\sigma }n_{\sigma }X_{%
\overline{\sigma }}^{\mp })}(t_{1},t_{2})\right]  \notag
\end{eqnarray}%
where, the indexes $n,j$ of the correlation functions $F_{niljq\sigma
}^{(n_{\sigma }X_{\overline{\sigma }}^{\pm })}(t_{1},t_{2})$ are not equal
to each other, $n\neq j$. In the above EOM there also emerge new correlation
functions with $L=2$, $F_{nmiljq\sigma }^{(X_{\sigma }^{-}X_{\overline{%
\sigma }}^{\pm })}(t_{1},t_{2})$, as well as some ones with $L=3$, $%
F_{nilkjq\sigma }^{(n_{\sigma }n_{\sigma }X_{\overline{\sigma }}^{\mp
})}(t_{1},t_{2})$. Of course, following this same routine, we can write out
the EOMs of correlation functions with $L=3$ that appear in the above
equations, and so on. Meanwhile, there emerge a series of static correlation
functions in these hierarchical EOMs, that are corresponding to the equal
time parts of these EOMs, respectively, such as, $<\widehat{n}_{i\overline{%
\sigma }}>$, $<\widehat{X}_{il\overline{\sigma }}^{(\pm )}>$, $<\{\widehat{n}%
_{n\sigma }\widehat{c}_{j\sigma },\widehat{c}_{q\sigma }^{\dagger }\}%
\widehat{X}_{il\overline{\sigma }}^{\pm }>$, etc.. These static correlation
functions can be used to describe a variety of possible inhomogeneous states
of the system in low temperature regime, and they will strongly influence
the spectral density function of electrons.

More importantly, in the above calculations, it is clearly shown that in the
EOM of a multipe-site correlation function with $L$ there never emerges any
multipe-site correlation function belonging to the level $L-1$. It is an
important guideline as taking any cut-off approximation for a multipe-site
correlation function belonging to the high level $L+1$.

This set of the EOMs has an obvious hierarchical structure denoted by the
parameter $L$. In the EOM of the one-particle Green function $G_{lj\sigma
}(\omega )$, there only emerges the correlation function $F_{iij\sigma
}^{(n_{\overline{\sigma }})}(\omega )$ which is produced by the on-site
Coulomb interaction. In the EOM of the correlation function $F_{ilj\sigma
}^{(n_{\overline{\sigma }})}(\omega )$, there appear the correlation
functions $F_{illj\sigma }^{(n_{\overline{\sigma }}n_{\overline{\sigma }%
})}(\omega )$ and $F_{iklj\sigma }^{(X_{\overline{\sigma }}^{-})}(\omega )$.
The correlation function $F_{illj\sigma }^{(n_{\overline{\sigma }}n_{%
\overline{\sigma }})}(\omega )$ originates from the on-site Coulomb
interaction, which can be written as, $F_{illj\sigma }^{(n_{\overline{\sigma 
}}n_{\overline{\sigma }})}(\omega )=\delta _{il}F_{ilj\sigma }^{(n_{%
\overline{\sigma }})}(\omega )+\left( 1-\delta _{il}\right) F_{illj\sigma
}^{(n_{\overline{\sigma }}n_{\overline{\sigma }})}(\omega )$, and the
correlation function $\left( 1-\delta _{il}\right) F_{illj\sigma }^{(n_{%
\overline{\sigma }}n_{\overline{\sigma }})}(\omega )$ is expectantly small
as assuming the function $F_{ilj\sigma }^{(n_{\overline{\sigma }})}(\omega )$
varying smoothly in the lattice space; Thus in this way, the on-site Coulomb
interaction of electrons is rigorously treated. While the correlation
function $F_{iklj\sigma }^{(X_{\overline{\sigma }}^{-})}(\omega )$ comes
from the kinetic energy part of the Hamiltonian. Thus the correlation
function $F_{ilj\sigma }^{(n_{\overline{\sigma }})}(\omega )$ in fact
represents the correlation effect of electrons from the mixture of both the
kinetic energy and interaction potential parts of the system.

In the thermodynamic limit, this set of the hierarchical EOMs cannot be
automatically closed. To solving this set of equations, for simplicity, we
would cut off the EOMs at the level $L=2$, thus we must take cut-off
approximations for those multipe-site correlation functions belonging to the
level $L=3$ in these EOMs. According to the hierarchical character of the
EOM of a multipe-site correlation function, we take the following cut-off
approximations for the correlation functions belonging to the level $L=3$ in
the Eq.(\ref{13a}),%
\begin{eqnarray}
F_{niljjq\sigma }^{(n_{\sigma }X_{\overline{\sigma }}^{\pm }n_{\overline{%
\sigma }})}(t_{1},t_{2})\simeq < &&\widehat{n}_{j\overline{\sigma }%
}(t_{1})>F_{niljq\sigma }^{(n_{\sigma }X_{\overline{\sigma }}^{\pm
})}(t_{1},t_{2}),\text{ }j\neq i,l  \notag \\
F_{nliljq\sigma }^{(n_{\sigma }n_{\sigma }X_{\overline{\sigma }}^{\mp
})}(t_{1},t_{2})\simeq < &&\widehat{n}_{n\sigma }(t_{1})>F_{liljq\sigma
}^{(n_{\sigma }X_{\overline{\sigma }}^{\mp })}(t_{1},t_{2})  \label{13c} \\
+ &<&\widehat{n}_{l\sigma }(t_{1})>F_{niljq\sigma }^{(n_{\sigma }X_{%
\overline{\sigma }}^{\mp })}(t_{1},t_{2}),\text{ }n\neq l;\text{ }j\neq n,l 
\notag
\end{eqnarray}%
Under these cut-off approximations, the subset of the EOMs of the
correlation functions $F_{niljq\sigma }^{(n_{\sigma }X_{\overline{\sigma }%
}^{\pm })}(t_{1},t_{2})$ is closed after taking, $F_{nmiljq\sigma
}^{(X_{\sigma }^{-}X_{\overline{\sigma }}^{\pm })}(t_{1},t_{2})\rightarrow 0$%
, where it is assumed that they have small contribution to the correlation
functions $F_{niljq\sigma }^{(n_{\sigma }X_{\overline{\sigma }}^{\pm
})}(t_{1},t_{2})$. These cut-off approximations are consistent with the
hierarchical character of the EOM of a multipe-site correlation function
where any multipe-site correlation function with $L-1$ does not appear in
the subset of the EOMs of the multipe-site correlation functions with $L$.
On the other hand, these cut-off approximations are qualitatively distinct
from usual ones taken in perturbation theories, such as usual Hartree-Fock
approximation and/or random-phase approximation (RPA), because before taking
the above cut-off approximations, the on-site Coulomb interaction of
electrons has been rigorously treated in the subset of the EOMs of the
correlation functions with $L=1$. That is, before taking any approximation,
the Coulomb interaction of electrons has been effectively treated rather
than taking the Coulomb interaction potential as a perturbative term like
that doing in usual perturbation theories.

After taking the above approximations, the EOMs of the correlation functions 
$F_{iljq\sigma }^{(X_{\overline{\sigma }}^{\pm })}(t_{1},t_{2})$ and $%
F_{niljq\sigma }^{(n_{\sigma }X_{\overline{\sigma }}^{\pm })}(t_{1},t_{2})$
can be further simplified by defining the following new functions,%
\begin{eqnarray}
F_{iljq\sigma }^{(\pm X_{\overline{\sigma }})}(t_{1},t_{2}) &=&\frac{1}{2}%
\left[ F_{iljq\sigma }^{(X_{\overline{\sigma }}^{+})}(t_{1},t_{2})\pm
F_{iljq\sigma }^{(X_{\overline{\sigma }}^{-})}(t_{1},t_{2})\right]  \notag \\
F_{niljq\sigma }^{(\pm n_{\sigma }X_{\overline{\sigma }})}(t_{1},t_{2}) &=&%
\frac{1}{2}\left[ F_{niljq\sigma }^{(n_{\sigma }X_{\overline{\sigma }%
}^{+})}(t_{1},t_{2})\pm F_{niljq\sigma }^{(n_{\sigma }X_{\overline{\sigma }%
}^{-})}(t_{1},t_{2})\right]  \label{14}
\end{eqnarray}%
which can be used to decouple the EOMs of the correlation functions $%
F_{iljq\sigma }^{(X_{\overline{\sigma }}^{\pm })}(t_{1},t_{2})$ and $%
F_{niljq\sigma }^{(n_{\sigma }X_{\overline{\sigma }}^{\pm })}(t_{1},t_{2})$,
respectively. With these new functions, the Eq.(\ref{13a}) can be rewritten
as that, 
\begin{eqnarray}
\left[ i\partial _{t_{1}}-U_{nilj}^{+}\right] F_{niljq\sigma }^{(+n_{\sigma
}X_{\overline{\sigma }})}(t_{1},t_{2}) &=&\delta (t_{1}-t_{2})<\widehat{c}_{i%
\overline{\sigma }}^{\dagger }\widehat{c}_{l\overline{\sigma }}\left[ 
\widehat{n}_{n\sigma }\delta _{jq}-\widehat{c}_{q\sigma }^{\dagger }\widehat{%
c}_{j\sigma }\delta _{nq}\right] >  \notag \\
&&+\sum_{m}\left( \widehat{h}_{jm}-\mu \delta _{jm}\right) F_{nilmq\sigma
}^{(+n_{\sigma }X_{\overline{\sigma }})}(t_{1},t_{2})  \label{15a} \\
&&-\sum_{m}\left[ \widehat{h}_{im}F_{nmljq\sigma }^{(+n_{\sigma }X_{%
\overline{\sigma }})}(t_{1},t_{2})-\widehat{h}_{lm}F_{nimjq\sigma
}^{(+n_{\sigma }X_{\overline{\sigma }})}(t_{1},t_{2})\right]  \notag
\end{eqnarray}%
\begin{eqnarray}
\left[ i\partial _{t_{1}}-U_{nilj}^{-}\right] F_{niljq\sigma }^{(-n_{\sigma
}X_{\overline{\sigma }})}(t_{1},t_{2}) &=&\delta (t_{1}-t_{2})<\widehat{c}_{l%
\overline{\sigma }}^{\dagger }\widehat{c}_{i\overline{\sigma }}\left[ 
\widehat{n}_{n\sigma }\delta _{jq}-\widehat{c}_{q\sigma }^{\dagger }\widehat{%
c}_{j\sigma }\delta _{nq}\right] >  \notag \\
&&+\sum_{m}\left( \widehat{h}_{jm}-\mu \delta _{jm}\right) F_{nilmq\sigma
}^{(-n_{\sigma }X_{\overline{\sigma }})}(t_{1},t_{2})  \label{15b} \\
&&+\sum_{m}\left[ \widehat{h}_{im}F_{nmljq\sigma }^{(-n_{\sigma }X_{%
\overline{\sigma }})}(t_{1},t_{2})-\widehat{h}_{lm}F_{nimjq\sigma
}^{(-n_{\sigma }X_{\overline{\sigma }})}(t_{1},t_{2})\right]  \notag
\end{eqnarray}%
where $U_{nilj}^{\pm }=\Delta _{nilj}^{\pm }+U\left( \delta _{lj}\pm \delta
_{nl}\mp \delta _{ni}\right) $, and $\Delta _{nilj}^{\pm }=U\left[ \left(
1-\delta _{ij}\right) \left( 1-\delta _{lj}\right) n_{\overline{\sigma }}\pm
\left( \delta _{ni}-\delta _{nl}\right) n_{\sigma }\right] $ which is
contributed by the correlation functions belonging to the level $L=3$ under
the cut-off approximations in Eq.(\ref{13c}). The Eq.(\ref{13}) can be
rewritten as that,%
\begin{eqnarray}
\left[ i\partial _{t_{1}}-U\delta _{lj}\right] F_{iljq\sigma }^{(+X_{%
\overline{\sigma }})}(t_{1},t_{2}) &=&\delta (t_{1}-t_{2})<\widehat{c}_{i%
\overline{\sigma }}^{\dagger }\widehat{c}_{l\overline{\sigma }}>\delta _{jq}
\notag \\
&&+U\left( \delta _{nl}-\delta _{ni}\right) F_{niljq\sigma }^{(+n_{\sigma
}X_{\overline{\sigma }})}(t_{1},t_{2})  \notag \\
&&+\sum_{m}\left( \widehat{h}_{jm}-\mu \delta _{jm}\right) F_{ilmq\sigma
}^{(+X_{\overline{\sigma }})}(t_{1},t_{2})  \label{16a} \\
&&-\sum_{m}\left[ \widehat{h}_{im}F_{mljq\sigma }^{(+X_{\overline{\sigma }%
})}(t_{1},t_{2})-\widehat{h}_{lm}F_{imjq\sigma }^{(+X_{\overline{\sigma }%
})}(t_{1},t_{2})\right]  \notag
\end{eqnarray}%
\begin{eqnarray}
\left[ i\partial _{t_{1}}-U\delta _{ij}\right] F_{iljq\sigma }^{(-X_{%
\overline{\sigma }})}(t_{1},t_{2}) &=&\delta (t_{1}-t_{2})<\widehat{c}_{l%
\overline{\sigma }}^{\dagger }\widehat{c}_{i\overline{\sigma }}>\delta _{jq}
\notag \\
&&+U\left( \delta _{ni}-\delta _{nl}\right) F_{niljq\sigma }^{(-n_{\sigma
}X_{\overline{\sigma }})}(t_{1},t_{2})  \notag \\
&&+\sum_{m}\left( \widehat{h}_{jm}-\mu \delta _{jm}\right) F_{ilmq\sigma
}^{(-X_{\overline{\sigma }})}(t_{1},t_{2})  \label{16b} \\
&&+\sum_{m}\left[ \widehat{h}_{im}F_{mljq\sigma }^{(-X_{\overline{\sigma }%
})}(t_{1},t_{2})-\widehat{h}_{lm}F_{imjq\sigma }^{(-X_{\overline{\sigma }%
})}(t_{1},t_{2})\right]  \notag
\end{eqnarray}%
where the contribution from the correlation functions $\left( 1-\delta
_{ij}\right) \left( 1-\delta _{lj}\right) F_{iljjq\sigma }^{(X_{\overline{%
\sigma }}^{\mp }n_{\overline{\sigma }})}(t_{1},t_{2})$ has been neglected,
which is assumed to be small. While the Eq.(\ref{12a}) can be rewritten as
that,%
\begin{eqnarray}
\left[ i\partial _{t_{1}}-U\delta _{ij}\right] F_{ijq\sigma }^{(n_{\overline{%
\sigma }})}(t_{1},t_{2}) &=&\delta (t_{1}-t_{2})<\widehat{n}_{i\overline{%
\sigma }}>\delta _{jq}  \notag \\
&&+\sum_{m}\left( \widehat{h}_{jm}-\mu \delta _{jm}\right) F_{imq\sigma
}^{(n_{\overline{\sigma }})}(t_{1},t_{2})  \label{17} \\
&&+\sum_{m}\widehat{h}_{im}\left[ F_{imjq\sigma }^{(+X_{\overline{\sigma }%
})}(t_{1},t_{2})-F_{imjq\sigma }^{(-X_{\overline{\sigma }})}(t_{1},t_{2})%
\right]  \notag
\end{eqnarray}%
where we have neglected the contribution coming from the function $\left(
1-\delta _{ij}\right) F_{ijjq\sigma }^{(n_{\overline{\sigma }}n_{\overline{%
\sigma }})}(t_{1},t_{2})$. As the correlation function $F_{ijjq\sigma }^{(n_{%
\overline{\sigma }}n_{\overline{\sigma }})}(t_{1},t_{2})$ is a smoothly
varying function in the lattice space, the contribution coming from the
function $\left( 1-\delta _{ij}\right) F_{ijjq\sigma }^{(n_{\overline{\sigma 
}}n_{\overline{\sigma }})}(t_{1},t_{2})$ can be neglected as a simple
approximation.

\section{Solution of the EOM}

The linear EOMs in Eqs.(\ref{15a}-\ref{17}) can be solved after taking the
Fourier representation of the multipe-site correlation functions, in which
one needs more carefully as taking the Fourier transformation because these
functions depend upon more than two lattice sites, i.e., they are the
correlation functions defined by more than two operators residing at
different sites. Moreover, the multipe-site functions $F_{ijq\sigma }^{(n_{%
\overline{\sigma }})}(t_{1},t_{2})$, $F_{imjq\sigma }^{(\pm X_{\overline{%
\sigma }})}(t_{1},t_{2})$ and $F_{niljq\sigma }^{(\pm n_{\sigma }X_{%
\overline{\sigma }})}(t_{1},t_{2})$ can be seen as "tensors" with different
ranks, respectively, and the above linear EOMs in fact are the equations of
the tensors.

With the help of the Fourier transformations of the electron operators,%
\begin{eqnarray}
\widehat{c}_{i\sigma }\left( t\right) &=&\frac{1}{\sqrt{N}}\sum_{\mathbf{k}}%
\widehat{c}_{k\sigma }\left( \omega \right) e^{i\omega t-i\mathbf{k\cdot x}%
_{i}}  \notag \\
\widehat{c}_{i\sigma }^{\dag }\left( t\right) &=&\frac{1}{\sqrt{N}}\sum_{%
\mathbf{k}}\widehat{c}_{k\sigma }^{\dag }\left( \omega \right) e^{i\mathbf{%
k\cdot x}_{i}-i\omega t}  \label{19}
\end{eqnarray}%
where the site $i$ is denoted by $\mathbf{x}_{i}$, we take the following
Fourier representations of the multipe-site correlation functions according
to their definition in Eq.(\ref{8}), respectively,%
\begin{eqnarray}
G_{iq\sigma }(t_{1},t_{2}) &=&\frac{1}{N}\sum_{\mathbf{k}}G_{k\sigma
}(\omega ;iq)e^{i\omega \left( t_{1}-t_{2}\right) -i\mathbf{k\cdot x}_{iq}} 
\notag \\
F_{ijq\sigma }^{(n_{\overline{\sigma }})}(t_{1},t_{2}) &=&\frac{1}{N^{2}}%
\sum_{\mathbf{kk}^{\prime }}F_{k^{\prime }k\overline{\sigma }\sigma }^{(n_{%
\overline{\sigma }})}(\omega ;ijq)e^{i\mathbf{k}^{\prime }\mathbf{\cdot x}%
_{ij}}e^{i\omega \left( t_{1}-t_{2}\right) -i\mathbf{k\cdot x}_{jq}}  \notag
\\
F_{iljq\sigma }^{(+X_{\overline{\sigma }})}(t_{1},t_{2}) &=&\frac{1}{N^{2}}%
\sum_{\mathbf{kk}^{\prime }\mathbf{k}_{1}}F_{k_{1}k^{\prime }k\overline{%
\sigma }\sigma }^{(+X_{\overline{\sigma }})}(\omega ;iljq)e^{i\mathbf{k}_{1}%
\mathbf{\cdot x}_{ij}-i\mathbf{k}^{\prime }\mathbf{\cdot x}_{lj}}e^{i\omega
\left( t_{1}-t_{2}\right) -i\mathbf{k\cdot x}_{jq}}  \notag \\
F_{iljq\sigma }^{(-X_{\overline{\sigma }})}(t_{1},t_{2}) &=&\frac{1}{N^{2}}%
\sum_{\mathbf{kk}^{\prime }\mathbf{k}_{1}}F_{k_{1}k^{\prime }k\overline{%
\sigma }\sigma }^{(-X_{\overline{\sigma }})}(\omega ;iljq)e^{i\mathbf{k}_{1}%
\mathbf{\cdot x}_{lj}-i\mathbf{k}^{\prime }\mathbf{\cdot x}_{ij}}e^{i\omega
\left( t_{1}-t_{2}\right) -i\mathbf{k\cdot x}_{jq}}  \label{20} \\
F_{niljq\sigma }^{(+n_{\sigma }X_{\overline{\sigma }})}(t_{1},t_{2}) &=&%
\frac{1}{N^{3}}\sum_{\mathbf{kp}^{\prime }\mathbf{k}^{\prime }\mathbf{k}%
_{1}}F_{p^{\prime }k_{1}k^{\prime }k\overline{\sigma }\sigma }^{(+n_{\sigma
}X_{\overline{\sigma }})}(\omega ;niljq)e^{i\mathbf{p}^{\prime }\mathbf{%
\cdot x}_{nj}}e^{i\mathbf{k}_{1}\mathbf{\cdot x}_{ij}-i\mathbf{k}^{\prime }%
\mathbf{\cdot x}_{lj}}e^{i\omega \left( t_{1}-t_{2}\right) -i\mathbf{k\cdot x%
}_{jq}}  \notag \\
F_{niljq\sigma }^{(-n_{\sigma }X_{\overline{\sigma }})}(t_{1},t_{2}) &=&%
\frac{1}{N^{3}}\sum_{\mathbf{kp}^{\prime }\mathbf{k}^{\prime }\mathbf{k}%
_{1}}F_{p^{\prime }k_{1}k^{\prime }k\overline{\sigma }\sigma }^{(-n_{\sigma
}X_{\overline{\sigma }})}(\omega ;niljq)e^{i\mathbf{p}^{\prime }\mathbf{%
\cdot x}_{nj}}e^{i\mathbf{k}_{1}\mathbf{\cdot x}_{lj}-i\mathbf{k}^{\prime }%
\mathbf{\cdot x}_{ij}}e^{i\omega \left( t_{1}-t_{2}\right) -i\mathbf{k\cdot x%
}_{jq}}  \notag
\end{eqnarray}%
where the "Fourier functions" $G_{k\sigma }(\omega ;iq)$, $F_{k^{\prime }k%
\overline{\sigma }\sigma }^{(n_{\overline{\sigma }})}(\omega ;ijq)$, $%
F_{k_{1}k^{\prime }k\overline{\sigma }\sigma }^{(\pm X_{\overline{\sigma }%
})}(\omega ;iljq)$ and $F_{p^{\prime }k_{1}k^{\prime }k\overline{\sigma }%
\sigma }^{(\pm n_{\sigma }X_{\overline{\sigma }})}(\omega ;niljq)$ generally
depend upon the lattice sites, because the corresponding correlation
functions depend upon more than two lattice sites. However, the function $%
G_{k\sigma }(\omega ;iq)$ depends upon the lattice site due to in its EOM
there appears the function $F_{k^{\prime }k\overline{\sigma }\sigma }^{(n_{%
\overline{\sigma }})}(\omega ;iiq)$.

Using the Fourier representations in Eq.(\ref{20}), the EOMs in Eqs.(\ref%
{15a}-\ref{17}) become a set of linear algebraic equations of the Fourier
functions, in which the terms with the operator $\widehat{h}_{im}$ in the
right hand side of these EOMs make this set of algebraic equations be hardly
solved because they depend upon the values of these functions on the nearest
neighbour sites. If we assume that the Fourier functions $G_{k\sigma
}(\omega ;iq)$, $F_{k^{\prime }k\overline{\sigma }\sigma }^{(n_{\overline{%
\sigma }})}(\omega ;ijq)$, $F_{k_{1}k^{\prime }k\overline{\sigma }\sigma
}^{(\pm X_{\overline{\sigma }})}(\omega ;iljq)$ and $F_{p^{\prime
}k_{1}k^{\prime }k\overline{\sigma }\sigma }^{(\pm n_{\sigma }X_{\overline{%
\sigma }})}(\omega ;niljq)$ are smoothly and slowly varying functions on the
lattice sites, the difference of a Fourier function between two nearest
neighbour sites would be a small quantity, and it can be safely neglected.
Therefore, we can further take the following approximation as in solving
these EOMs,%
\begin{equation}
\sum_{m}\widehat{h}_{jm}X_{\left\{ pk\right\} \overline{\sigma }\sigma
}(\omega ;\left\{ imq\right\} )\simeq \sum_{m}\widehat{h}_{jm}X_{\left\{
pk\right\} \overline{\sigma }\sigma }(\omega ;\left\{ ijq\right\} )
\label{21}
\end{equation}%
where $X_{\left\{ pk\right\} \overline{\sigma }\sigma }(\omega ;\left\{
ijq\right\} )=G_{k\sigma }(\omega ;iq)$, $F_{k^{\prime }k\overline{\sigma }%
\sigma }^{(n_{\overline{\sigma }})}(\omega ;ijq)$, $F_{k_{1}k^{\prime }k%
\overline{\sigma }\sigma }^{(\pm X_{\overline{\sigma }})}(\omega ;iljq)$, $%
F_{p^{\prime }k_{1}k^{\prime }k\overline{\sigma }\sigma }^{(\pm n_{\sigma
}X_{\overline{\sigma }})}(\omega ;niljq)$.

Under the approximation in Eq.(\ref{21}), the EOMs in Eqs.(\ref{15a}-\ref{17}%
) can be analytically solved. For simplicity, the one-particle Green
function $G_{k\sigma }(\omega )$ can be written as following three parts,%
\begin{equation}
G_{k\sigma }(\omega )=G_{k\sigma }^{S}(\omega )+G_{k\sigma }^{LC}(\omega
)+G_{k\sigma }^{NLC}(\omega )  \label{22}
\end{equation}%
where the Green function $G_{k\sigma }^{S}(\omega )$ is contributed by the
functions $F_{k^{\prime }k\overline{\sigma }\sigma }^{(n_{\overline{\sigma }%
})}(\omega ;ijq)$ and $F_{k_{1}k^{\prime }k\overline{\sigma }\sigma }^{(\pm
X_{\overline{\sigma }})}(\omega ;iljq)$ without considering the functions $%
F_{p^{\prime }k_{1}k^{\prime }k\overline{\sigma }\sigma }^{(\pm n_{\sigma
}X_{\overline{\sigma }})}(\omega ;niljq)$, and the Green function $%
G_{k\sigma }^{LC}(\omega )$ and $G_{k\sigma }^{NLC}(\omega )$ are
contributed by the functions $F_{p^{\prime }k_{1}k^{\prime }k\overline{%
\sigma }\sigma }^{(\pm n_{\sigma }X_{\overline{\sigma }})}(\omega ;niljq)$,
in which there naturally emerge the static correlation functions $%
N_{k_{1}k^{\prime }}=<\widehat{n}_{k_{1}\sigma }\widehat{c}_{k^{\prime
}+k_{1}\overline{\sigma }}^{\dag }\widehat{c}_{k^{\prime }\overline{\sigma }%
}>$ and $C_{pk_{1}k^{\prime }k}=<\widehat{c}_{k+p\sigma }^{\dagger }\widehat{%
c}_{k+p+k_{1}-k^{\prime }\sigma }\widehat{c}_{k_{1}\overline{\sigma }}^{\dag
}\widehat{c}_{k^{\prime }\overline{\sigma }}>$. The appearance of the static
correlation functions $N_{k_{1}k^{\prime }}$ and $C_{pk_{1}k^{\prime }k}$ in
the one-particle Green function means that the possible intertwined orders
in low temperature regime derived by the strong Coulomb interaction would
drastically influence the spectral density function of electeons. On the
other hand, it is also shown that the HGFA can uncover the intrinsic
relation between the one-particle Green function and a variety of
inhomogeneous states in highly correlated systems.

The Green function $G_{k\sigma }^{S}(\omega )$ can be written as that, 
\begin{eqnarray}
G_{k\sigma }^{S}(\omega ) &=&\left[ 1-n_{\overline{\sigma }}+\frac{1}{N}%
\sum_{k_{1}}\left( \frac{\xi _{k_{1}}n_{k_{1}\overline{\sigma }}}{\epsilon
_{k_{1}}}-\frac{\xi _{k_{1}}n_{k_{1}\overline{\sigma }}}{\epsilon _{k_{1}}+U}%
\right) \right] \frac{1}{\omega -\varepsilon _{k}}  \notag \\
&&+\left[ n_{\overline{\sigma }}+\frac{1}{N}\sum_{k_{1}}\left( \frac{\xi
_{k_{1}}n_{k_{1}\overline{\sigma }}}{U-\epsilon _{k_{1}}}+\frac{\xi
_{k_{1}}n_{k_{1}\overline{\sigma }}}{\epsilon _{k_{1}}}\right) \right] \frac{%
1}{\omega -\varepsilon _{k}-U}  \notag \\
&&-\frac{1}{N}\sum_{k_{1}}\left( \frac{\xi _{k_{1}}n_{k_{1}\overline{\sigma }%
}}{\epsilon _{k_{1}}}+\frac{\xi _{k_{1}}n_{k_{1}\overline{\sigma }}}{%
U-\epsilon _{k_{1}}}\right) \frac{1}{\omega -\varepsilon _{k}-\epsilon
_{k_{1}}}  \label{23a} \\
&&-\frac{1}{N}\sum_{k_{1}}\left( \frac{\xi _{k_{1}}n_{k_{1}\overline{\sigma }%
}}{\epsilon _{k_{1}}}-\frac{\xi _{k_{1}}n_{k_{1}\overline{\sigma }}}{%
\epsilon _{k_{1}}+U}\right) \frac{1}{\omega -\varepsilon _{k}-\epsilon
_{k_{1}}-U}  \notag
\end{eqnarray}%
where $n_{\overline{\sigma }}=<\widehat{n}_{i\overline{\sigma }}>$, $n_{k%
\overline{\sigma }}=<\widehat{c}_{k\overline{\sigma }}^{\dag }\widehat{c}_{k%
\overline{\sigma }}>$, $\varepsilon _{k}=\xi _{k}-\mu $, $\epsilon
_{k}=4t_{0}+\xi _{k}$, and $\xi _{k}=-2t_{0}\left[ \cos \left( ak_{x}\right)
+\cos \left( ak_{y}\right) \right] $, here $a$ is the lattice constant.
Notice that the Green function $G_{k\sigma }^{S}(\omega )$ has a similar
expression like that for an one site Hubbard model, in which as taking, $%
t_{0}=0$, it is reduced to a two level structure form of the one-particle
Green function for one site Hubbard model. There is a gap between the lower
and upper Hubbard bands, which is proportional to $U$ in the large $U$
limit. However, the appearence of the third term in the right hand side
reveals that there emerge new states within this gap because the function $%
\epsilon _{k}$ is positive, $\epsilon _{k}\geq 0$.

The Green function $G_{k\sigma }^{LC}(\omega )$ reads that,%
\begin{eqnarray}
G_{k\sigma }^{LC}(\omega ) &=&-\frac{1}{N^{2}}\sum_{k^{\prime }k_{1}}\frac{%
A_{k_{1}k^{\prime }}^{L}N_{k_{1}k^{\prime }}}{1-n_{\sigma }}\frac{1}{\omega
-\varepsilon _{k}}+\frac{1}{N^{2}}\sum_{k^{\prime }k_{1}}\frac{%
B_{k_{1}k^{\prime }}^{L}N_{k_{1}k^{\prime }}}{1-n_{\sigma }}\frac{1}{\omega
-\varepsilon _{k}-U}  \notag \\
&&-\frac{1}{N^{2}}\sum_{k^{\prime }k_{1}}\left( \frac{\xi _{k_{1}+k^{\prime
}}}{\eta _{k_{1}k^{\prime }}}-\frac{\xi _{k_{1}+k^{\prime }}}{\eta
_{k_{1}k^{\prime }}+U}\right) \frac{N_{k_{1}k^{\prime }}\left( 1-n_{\sigma
}\right) ^{-1}}{\omega -E_{k}\left( k_{1},k^{\prime }\right) }  \notag \\
&&+\frac{1}{N^{2}}\sum_{k^{\prime }k_{1}}\left( \frac{\xi _{k_{1}+k^{\prime
}}}{\eta _{k_{1}k^{\prime }}-\left( 1-n_{\sigma }\right) U}-\frac{\xi
_{k_{1}+k^{\prime }}}{\eta _{k_{1}k^{\prime }}+n_{\sigma }U}\right) \frac{%
N_{k_{1}k^{\prime }}\left( 1-n_{\sigma }\right) ^{-1}}{\omega -E_{k}\left(
k_{1},k^{\prime }\right) -\left( 1-n_{\sigma }\right) U}  \notag \\
&&+\frac{1}{N^{2}}\sum_{k^{\prime }k_{1}}\left( \frac{\xi _{k^{\prime }}}{%
\eta _{k_{1}k^{\prime }}-U}-\frac{\xi _{k^{\prime }}}{\eta _{k_{1}k^{\prime
}}}\right) \frac{N_{k_{1}k^{\prime }}\left( 1-n_{\sigma }\right) ^{-1}}{%
\omega -E_{k}\left( k_{1},k^{\prime }\right) -U}  \label{23b} \\
&&-\frac{1}{N^{2}}\sum_{k^{\prime }k_{1}}\left( \frac{\xi _{k^{\prime }}}{%
\eta _{k_{1}k^{\prime }}-n_{\sigma }U}-\frac{\xi _{k^{\prime }}}{\eta
_{k_{1}k^{\prime }}+\left( 1-n_{\sigma }\right) U}\right) \frac{%
N_{k_{1}k^{\prime }}\left( 1-n_{\sigma }\right) ^{-1}}{\omega -E_{k}\left(
k_{1},k^{\prime }\right) -n_{\sigma }U}  \notag
\end{eqnarray}%
where $\widehat{n}_{q\overline{\sigma }}=\sum_{\mathbf{k}}\widehat{c}_{k%
\overline{\sigma }}^{\dag }\widehat{c}_{k+q\overline{\sigma }}$, $%
E_{k}\left( k_{1},k^{\prime }\right) =\varepsilon _{k}-\eta _{k_{1}k^{\prime
}}$, $\eta _{k_{1}k^{\prime }}=\varepsilon _{k_{1}}-\varepsilon _{k^{\prime
}}$, $A_{k_{1}k^{\prime }}^{L}=-\frac{\xi _{k_{1}+k^{\prime }}}{\eta
_{k_{1}k^{\prime }}}+\frac{\xi _{k_{1}+k^{\prime }}}{\eta _{k_{1}k^{\prime
}}-\left( 1-n_{\sigma }\right) U}+\frac{\xi _{k^{\prime }}}{\eta
_{k_{1}k^{\prime }}-U}-\frac{\xi _{k^{\prime }}}{\eta _{k_{1}k^{\prime
}}-n_{\sigma }U}$ and $B_{k_{1}k^{\prime }}^{L}=-\frac{\xi _{k_{1}+k^{\prime
}}}{\eta _{k_{1}k^{\prime }}+U}+\frac{\xi _{k_{1}+k^{\prime }}}{\eta
_{k_{1}k^{\prime }}+n_{\sigma }U}+\frac{\xi _{k^{\prime }}}{\eta
_{k_{1}k^{\prime }}}-\frac{\xi _{k^{\prime }}}{\eta _{k_{1}k^{\prime
}}+\left( 1-n_{\sigma }\right) U}$. The Green function $G_{k\sigma
}^{LC}(\omega )$ is contributed by the static correlation function $%
N_{k_{1}k^{\prime }}=<\widehat{n}_{k_{1}\sigma }\widehat{c}_{k^{\prime
}+k_{1}\overline{\sigma }}^{\dag }\widehat{c}_{k^{\prime }\overline{\sigma }%
}>$, which is mainly related to the density-density correlation effect of
electrons. In contrast with the Green function $G_{k\sigma }^{S}(\omega )$,
there emerge new terms with poles residing around in the middle of the gap
between the lower and upper Hubbard bands in the large $U$ limit, and it
clearly reveals that there emerge new states in this regime. As taking a
simple Hartree-Fock approximation to the function $N_{k_{1}k^{\prime }}$, $%
N_{k_{1}k^{\prime }}\Rightarrow Nn_{\sigma }n_{k^{\prime }\overline{\sigma }%
}\delta _{0k_{1}}$, the Green function $G_{k\sigma }^{LC}(\omega )$ can be
incorporated into the Green function $G_{k\sigma }^{S}(\omega )$ to more
clearly show the emerged states in this gap.

The Green function $G_{k\sigma }^{NLC}(\omega )$ can be written as that,%
\begin{eqnarray}
G_{k\sigma }^{NLC}(\omega ) &=&\frac{1}{N^{2}}\sum_{pk^{\prime }k_{1}}\frac{%
A_{k_{1}k^{\prime }}^{NL}\left( p\right) C_{pk_{1}k^{\prime }k}}{1-n_{\sigma
}}\frac{1}{\omega -\varepsilon _{k+p}}-\frac{1}{N^{2}}\sum_{pk^{\prime
}k_{1}}\frac{B_{k_{1}k^{\prime }}^{NL}\left( p\right) C_{pk_{1}k^{\prime }k}%
}{1-n_{\sigma }}\frac{1}{\omega -\varepsilon _{k+p}-U}  \notag \\
&&+\frac{1}{N^{2}}\sum_{pk^{\prime }k_{1}}\left( \frac{\xi _{p-k^{\prime }}}{%
\eta _{k_{1}k^{\prime }}}-\frac{\xi _{p-k^{\prime }}}{\eta _{k_{1}k^{\prime
}}+U}\right) \frac{C_{pk_{1}k^{\prime }k}\left( 1-n_{\sigma }\right) ^{-1}}{%
\omega -E_{k+p}\left( k_{1},k^{\prime }\right) }  \notag \\
&&-\frac{1}{N^{2}}\sum_{pk^{\prime }k_{1}}\frac{U\xi _{p-k^{\prime }}}{%
\left( \eta _{k_{1}k^{\prime }}-\left( 1-n_{\sigma }\right) U\right) \left(
\eta _{k_{1}k^{\prime }}+n_{\sigma }U\right) }\frac{C_{pk_{1}k^{\prime
}k}\left( 1-n_{\sigma }\right) ^{-1}}{\omega -E_{k+p}\left( k_{1},k^{\prime
}\right) -\left( 1-n_{\sigma }\right) U}  \notag \\
&&-\frac{1}{N^{2}}\sum_{pk^{\prime }k_{1}}\left( \frac{\xi _{p+k_{1}}}{\eta
_{k_{1}k^{\prime }}-U}-\frac{\xi _{p+k_{1}}}{\eta _{k_{1}k^{\prime }}}%
\right) \frac{C_{pk_{1}k^{\prime }k}\left( 1-n_{\sigma }\right) ^{-1}}{%
\omega -E_{k+p}\left( k_{1},k^{\prime }\right) -U}  \label{23c} \\
&&+\frac{1}{N^{2}}\sum_{pk^{\prime }k_{1}}\frac{U\xi _{p+k_{1}}}{\left( \eta
_{k_{1}k^{\prime }}-n_{\sigma }U\right) \left( \eta _{k_{1}k^{\prime
}}+\left( 1-n_{\sigma }\right) U\right) }\frac{C_{pk_{1}k^{\prime }k}\left(
1-n_{\sigma }\right) ^{-1}}{\omega -E_{k+p}\left( k_{1},k^{\prime }\right)
-n_{\sigma }U}  \notag
\end{eqnarray}%
where $A_{k_{1}k^{\prime }}^{NL}\left( p\right) =-\frac{\xi _{p-k^{\prime }}%
}{\eta _{k_{1}k^{\prime }}}+\frac{\xi _{p-k^{\prime }}}{\eta
_{k_{1}k^{\prime }}-\left( 1-n_{\sigma }\right) U}+\frac{\xi _{p+k_{1}}}{%
\eta _{k_{1}k^{\prime }}-U}-\frac{\xi _{p+k_{1}}}{\eta _{k_{1}k^{\prime
}}-n_{\sigma }U}$, and $B_{k_{1}k^{\prime }}^{NL}\left( p\right) =-\frac{\xi
_{p-k^{\prime }}}{\eta _{k_{1}k^{\prime }}+U}+\frac{\xi _{p-k^{\prime }}}{%
\eta _{k_{1}k^{\prime }}+n_{\sigma }U}+\frac{\xi _{p+k_{1}}}{\eta
_{k_{1}k^{\prime }}}-\frac{\xi _{p+k_{1}}}{\eta _{k_{1}k^{\prime }}+\left(
1-n_{\sigma }\right) U}$. The Green function $G_{k\sigma }^{NLC}(\omega )$
is contributed by the static correlation function $C_{pk_{1}k^{\prime }k}=<%
\widehat{c}_{k+p\sigma }^{\dagger }\widehat{c}_{k+p+k_{1}-k^{\prime }\sigma }%
\widehat{c}_{k_{1}\overline{\sigma }}^{\dag }\widehat{c}_{k^{\prime }%
\overline{\sigma }}>$, and it is mainly related to the spin-spin correlation
effect of electrons and possible electronic pairing density correlation. On
the other hand, the function $C_{pk_{1}k^{\prime }k}$ makes the Fourier
function $G_{k\sigma }(\omega ;iq)$ depend upon the lattice coordinate $%
\mathbf{x}_{i}-\mathbf{x}_{q}$, meanwhile it also produces new terms with
poles residing around in the middle of the gap between the lower and upper
Hubbard bands in the large $U$ limit, which shows that there emerge new
states in this regime.

It becomes more clear that the emerged states in the gap between the lower
and upper Hubbard bands in the large $U$ limit are coming from two aspects:
one is contributed by the static quantities $<\widehat{X}_{il\overline{%
\sigma }}^{(\pm )}>$, and another one is contributed by the static
correlation functions $<\{\widehat{n}_{n\sigma }\widehat{c}_{j\sigma },%
\widehat{c}_{q\sigma }^{\dagger }\}\widehat{X}_{il\overline{\sigma }}^{\pm
}> $, that are represented by the functions $N_{k_{1}k^{\prime }}$ and $%
C_{pk_{1}k^{\prime }k}$ in the phase space. However, the more high order
static correlation functions emerged in the series of the hierarchical EOMs
have expectantly small contribution, and they cannot drastically alter the
present calculations.

According to the expression of the one-particle Green function $G_{k\sigma
}(\omega )$ in Eq.(\ref{22}), we can demonstrate that the spectral density
function, $A_{\sigma }\left( \omega ,k\right) =-2\func{Im}G_{\sigma }(\omega
+0^{+},k)$, satisfies the following sum rule,%
\begin{equation}
\frac{1}{2\pi }\int_{-\infty }^{\infty }d\omega A_{\sigma }\left( \omega
,k\right) =1  \label{26}
\end{equation}%
which is independent of the approximations taken in solving the EOMs of the
multipe-site correlation functions. It is resulting in that the EOMs of the
multipe-site correlation functions are linear.

In order to show a more clear analytical expression of the spectral density
function $A_{\sigma }\left( \omega ,k\right) $, we need to take some
approximations for the static correlation functions $N_{k_{1}k^{\prime }}$
and $C_{pk_{1}k^{\prime }k}$ that emerge in the expressions of the Green
functions $G_{k\sigma }^{LC}(\omega )$ and $G_{k\sigma }^{NLC}(\omega )$ in
the Eqs.(\ref{23b}) and (\ref{23c}), respectively. For example, we can
rewrite out them as the following forms,%
\begin{eqnarray}
N_{k_{1}k^{\prime }}=< &&\widehat{n}_{k_{1}\sigma }\widehat{c}_{k^{\prime
}+k_{1}\overline{\sigma }}^{\dag }\widehat{c}_{k^{\prime }\overline{\sigma }%
}>_{o}+Nn_{\sigma }n_{k^{\prime }\overline{\sigma }}\delta _{0k_{1}}  \notag
\\
C_{pk_{1}k^{\prime }k}=< &&\widehat{c}_{k+p\sigma }^{\dagger }\widehat{c}%
_{k+p+k_{1}-k^{\prime }\sigma }\widehat{c}_{k_{1}\overline{\sigma }}^{\dag }%
\widehat{c}_{k^{\prime }\overline{\sigma }}>_{o}  \label{27} \\
&&+\Delta _{k_{1}}^{\ast }\Delta _{k^{\prime }}\delta
_{-k_{1}k+p}-S_{k^{\prime }}^{+}S_{k_{1}}^{-}\delta _{k+pk^{\prime }}  \notag
\end{eqnarray}%
where $\Delta _{k^{\prime }}=<\widehat{c}_{k^{\prime }\overline{\sigma }}%
\widehat{c}_{-k^{\prime }\sigma }>$ is an uniform electronic pairing order
parameter, $S_{k^{\prime }}^{+}=<\widehat{c}_{k^{\prime }\sigma }^{\dagger }%
\widehat{c}_{k^{\prime }\overline{\sigma }}>$ and $S_{k_{1}}^{-}=<\widehat{c}%
_{k_{1}\overline{\sigma }}^{\dag }\widehat{c}_{k_{1}\sigma }>$ are spin
order parameters. The fluctuation parts $<\widehat{n}_{k_{1}\sigma }\widehat{%
c}_{k^{\prime }+k_{1}\overline{\sigma }}^{\dag }\widehat{c}_{k^{\prime }%
\overline{\sigma }}>_{o}$ and $<\widehat{c}_{k+p\sigma }^{\dagger }\widehat{c%
}_{k+p+k_{1}-k^{\prime }\sigma }\widehat{c}_{k_{1}\overline{\sigma }}^{\dag }%
\widehat{c}_{k^{\prime }\overline{\sigma }}>_{o}$ can be corresponded to the
static correlation functions of the charge/spin density wave and the
electronic pairing density wave orders, respectively. For simplicity,
hereafter we do not consider these fluctuation parts. It would be pointed
out that the present approach only shows the influence of possible order
paramters, such as $\Delta _{k^{\prime }}$ and $S_{k^{\prime }}^{\pm }$,
etc., on the one-particle Green function, and it cannot answer how to fix
these order parameters. For example, the uniform electronic pairing
parameter $\Delta _{k\prime }$ and the spin order parameters $S_{k^{\prime
}}^{\pm }$ can be self-consistently determined by taking the minimun of the
ground state energy which can be calculated by the spectral density function 
$A_{\sigma }\left( \omega ,k\right) $ and the correlation function $%
F_{iii\sigma }^{(n_{\overline{\sigma }})}(t_{1},t_{1}+0^{+})$.

\section{The spectral density function}

With the help of the expressions of the Green functions $G_{k\sigma
}^{S}(\omega )$, $G_{k\sigma }^{LC}(\omega )$ and $G_{k\sigma }^{NLC}(\omega
)$ in Eqs.(\ref{23a}-\ref{23c}), under the approximations taken in Eqs.(\ref%
{27}), we can write out the analytical expression of the spectral density
function of electrons $A_{\sigma }\left( \omega ,k\right) $, in which there
exist two characteristic main peaks with the corresponding factors $\delta
\left( \omega -\varepsilon _{k}\right) $ and $\delta \left( \omega
-\varepsilon _{k}-U\right) $, respectively, that come from the lower and
upper Hubbard bands. The gap between these two main peaks is proportional to 
$U$ in the large $U$ limit. On the other hand, as the hole doping, there
emerge new states within this gap, and for enough large of the hole doping,
these emerged states would fill in this gap. It may be the origins of a
variety of novel low temperature behavior shown by the 2D square lattice
Hubbard model at the intermediate coupling $U$.

Under the approximations of the Eqs.(\ref{27}), and after discarding the
contributions from the fluctuation parts of the static correlation functions 
$N_{k_{1}k^{\prime }}$ and $C_{pk_{1}k^{\prime }k}$, we can write out the
spectral density function $A_{\sigma }\left( \omega ,k\right) $ as the
following three parts,%
\begin{equation}
A_{\sigma }\left( \omega ,k\right) =A_{\sigma }^{L-U}\left( \omega ,k\right)
+A_{\sigma }^{Em}\left( \omega ,k\right) +A_{\sigma }^{Con}\left( \omega
,k\right)  \label{28}
\end{equation}%
where the spectral density function $A_{\sigma }^{L-U}\left( \omega
,k\right) $ is the contribution coming from the lower and upper Hubbard
bands, the spectral density function $A_{\sigma }^{Em}\left( \omega
,k\right) $ is contributed by the emerged states of electrons due to the
doped holes, and the spectral density function $A_{\sigma }^{Con}\left(
\omega ,k\right) $ is related to the pairing order parameter $\Delta _{k}$
and spin order parameters $S_{k}^{\pm }$, which is contributed by the
emerged collective modes.

According to the one-particle Green function in Eq.(\ref{22}), the spectral
density function $A_{\sigma }^{L-U}\left( \omega ,k\right) $ can be written
as that,%
\begin{equation}
A_{\sigma }^{L-U}\left( \omega ,k\right) =2\pi \left[ 1-n_{\overline{\sigma }%
}-Z^{L}\right] \delta \left( \omega -\varepsilon _{k}\right) +2\pi \left[ n_{%
\overline{\sigma }}-Z^{U}\right] \delta \left( \omega -\varepsilon
_{k}-U\right)   \label{29}
\end{equation}%
where the constants $Z^{L}$ and $Z^{U}$ are that,%
\begin{eqnarray}
Z^{L} &=&-\frac{1}{N}\frac{1-2n_{\sigma }}{1-n_{\sigma }}\sum_{k_{1}}\left( 
\frac{\xi _{k_{1}}n_{k_{1}\overline{\sigma }}}{\epsilon _{k_{1}}}-\frac{\xi
_{k_{1}}n_{k_{1}\overline{\sigma }}}{\epsilon _{k_{1}}+U}\right)   \notag \\
&&+\frac{1}{N}\frac{n_{\sigma }}{1-n_{\sigma }}\sum_{k_{1}}\left( \frac{\xi
_{k_{1}}n_{k_{1}\overline{\sigma }}}{n_{\sigma }U+\epsilon _{k_{1}}}-\frac{%
\xi _{k_{1}}n_{k_{1}\overline{\sigma }}}{\epsilon _{k_{1}}+\left(
1-n_{\sigma }\right) U}\right)   \label{29a}
\end{eqnarray}%
\begin{eqnarray}
Z^{U} &=&-\frac{1}{N}\frac{1-2n_{\sigma }}{1-n_{\sigma }}\sum_{k_{1}}\left( 
\frac{\xi _{k_{1}}n_{k_{1}\overline{\sigma }}}{U-\epsilon _{k_{1}}}+\frac{%
\xi _{k_{1}}n_{k_{1}\overline{\sigma }}}{\epsilon _{k_{1}}}\right)   \notag
\\
&&-\frac{1}{N}\frac{n_{\sigma }}{1-n_{\sigma }}\sum_{k_{1}}\left( \frac{\xi
_{k_{1}}n_{k_{1}\overline{\sigma }}}{n_{\sigma }U-\epsilon _{k_{1}}}-\frac{%
\xi _{k_{1}}n_{k_{1}\overline{\sigma }}}{\left( 1-n_{\sigma }\right)
U-\epsilon _{k_{1}}}\right)   \label{29b}
\end{eqnarray}%
This expression of the spectral density function $A_{\sigma }^{L-U}\left(
\omega ,k\right) $ is obviously similar to that for one site case, and the
two peaks denoted respectively by the functions $\delta \left( \omega
-\varepsilon _{k}\right) $ and $\delta \left( \omega -\varepsilon
_{k}-U\right) $, corresponding to the lower and upper Hubbard bands, is
separated by $U$. At half filling, $n_{\sigma }=\frac{1}{2},\sigma =\uparrow
,\downarrow $, the constants $Z^{L}$ and $Z^{U}$ both are equal to zero, and
the spectral weight constants of electrons in both the lower and upper
Hubbard bands are the same, and equal $\frac{1}{2}$. Thus the system is a
Mott insulator with the gap proportional to $U$ in the large $U$ limit, in
which there exist some collective excitation modes (i.e., $A_{\sigma
}^{Em}\left( \omega ,k\right) =0$, $A_{\sigma }^{Con}\left( \omega ,k\right)
\neq 0$, see below). While, as the hole doping, $n_{\sigma }=\frac{1}{2}%
-\delta $, the constants $Z^{L}$ and $Z^{U}$ are positive, and proportional
to the hole doping concentration $\delta $ in the large $U$ limit. This
means that as the hole doping, the spectral weight of electrons in both the
lower and upper Hubbard bands is reduced, and the part of them is transfered
into the gap where there emerge new states.

The spectral density function $A_{\sigma }^{Em}\left( \omega ,k\right) $ can
be written as that, 
\begin{eqnarray}
A_{\sigma }^{Em}\left( \omega ,k\right)  &=&-\frac{2\pi }{N}\frac{%
1-2n_{\sigma }}{1-n_{\sigma }}\sum_{k_{1}}\frac{U\xi _{k_{1}}n_{k_{1}%
\overline{\sigma }}}{\epsilon _{k_{1}}\left( U-\epsilon _{k_{1}}\right) }%
\delta \left( \omega -\varepsilon _{k}-\epsilon _{k_{1}}\right)   \notag \\
&&-\frac{2\pi }{N}\frac{Un_{\sigma }}{1-n_{\sigma }}\sum_{k_{1}}\frac{\xi
_{k_{1}}n_{k_{1}\overline{\sigma }}\delta \left( \omega -\varepsilon
_{k}-\epsilon _{k_{1}}-\left( 1-n_{\sigma }\right) U\right) }{\left(
n_{\sigma }U-\epsilon _{k_{1}}\right) \left[ \epsilon _{k_{1}}+\left(
1-n_{\sigma }\right) U\right] }  \notag \\
&&+\frac{2\pi }{N}\frac{Un_{\sigma }}{1-n_{\sigma }}\sum_{k^{\prime }}\frac{%
\xi _{k_{1}}n_{k_{1}\overline{\sigma }}\delta \left( \omega -\varepsilon
_{k}-\epsilon _{k_{1}}-n_{\sigma }U\right) }{\left( \epsilon
_{k_{1}}+n_{\sigma }U\right) \left[ \left( 1-n_{\sigma }\right) U-\epsilon
_{k_{1}}\right] }  \label{30} \\
&&-\frac{2\pi }{N}\frac{1-2n_{\sigma }}{1-n_{\sigma }}\sum_{k_{1}}\frac{U\xi
_{k_{1}}n_{k_{1}\overline{\sigma }}}{\epsilon _{k_{1}}\left( \epsilon
_{k_{1}}+U\right) }\delta \left( \omega -\varepsilon _{k}-\epsilon
_{k_{1}}-U\right)   \notag
\end{eqnarray}%
It is noticeable that at the half filling $n_{\sigma }=\frac{1}{2}$, the
spectral density function $A_{\sigma }^{Em}\left( \omega ,k\right) $ is
zero, $A_{\sigma }^{Em}\left( \omega ,k\right) =0$. After the hole doping, $%
n_{\sigma }=\frac{1}{2}-\delta $, there emerge new states that reside in the
gap and at the top of the upper Hubbard bands, respectively. In the above
expression of the spectral density function $A_{\sigma }^{Em}\left( \omega
,k\right) $, the first three terms in the right hand side are contributed by
the emerged states in the gap, where the first term resides at the top of
the lower Hubbard band, which is proportional to the hole doping
concentration $\delta $. The second and third terms residing around in the
middle of the gap take converse values, and they are proportional to $\frac{1%
}{U}$ for large $U$. In the underdoped regime, after considering $A_{\sigma
}^{Con}\left( \omega ,k\right) $, the total spectral density function of
electrons $A_{\sigma }\left( \omega ,k\right) $ may show a dip around in the
middle of the gap. The last term in the right hand side is contributed by
the emerged states residing at the top of the upper Hubbard band, which is
also proportional to the hole doping concentration $\delta $. In the low
temperature limit, the first term of the spectral density function $%
A_{\sigma }^{Em}\left( \omega ,k\right) $ is mainly responsible for the
infrared reflectivity spectrum in the gap, where the effective carrier
density grows in proportion to the hole doping concentration $\delta $,
consistent with the experimental observations.

The spectral density function $A_{\sigma }^{Con}\left( \omega ,k\right) $
can be directly obtained by the Green function $G_{k\sigma }^{NLC}(\omega )$
which originates from the static correlation function $<\widehat{c}%
_{k+p\sigma }^{\dagger }\widehat{c}_{k+p+k_{1}-k^{\prime }\sigma }\widehat{c}%
_{k_{1}\overline{\sigma }}^{\dag }\widehat{c}_{k^{\prime }\overline{\sigma }%
}>$. Under the approximations in Eqs.(\ref{27}), it can be written as two
parts, 
\begin{equation}
A_{\sigma }^{Con}\left( \omega ,k\right) =A_{\sigma }^{spin}\left( \omega
,k\right) +A_{\sigma }^{pair}\left( \omega ,k\right)  \label{30a}
\end{equation}%
where the spectral density function $A_{\sigma }^{spin}\left( \omega
,k\right) $ is related to the spin parameters $S_{k^{\prime }}^{\pm }$,%
\begin{equation}
A_{\sigma }^{spin}\left( \omega ,k\right) =-\frac{2\pi }{\left( 1-n_{\sigma
}\right) N^{2}}\sum_{k^{\prime }k_{1}}\Gamma _{k_{1}k^{\prime }}\left(
\omega \right) S_{k^{\prime }}^{+}S_{k_{1}}^{-}  \label{31}
\end{equation}%
\begin{eqnarray}
\Gamma _{k_{1}k^{\prime }}\left( \omega \right) &=&A_{k_{1}k^{\prime
}}^{NL}\left( k^{\prime }-k\right) \delta \left( \omega -\varepsilon
_{k^{\prime }}\right) -B_{k_{1}k^{\prime }}^{NL}\left( k^{\prime }-k\right)
\delta \left( \omega -\varepsilon _{k^{\prime }}-U\right)  \notag \\
&&+\frac{U\xi _{-k}\delta \left( \omega +\eta _{k_{1}k^{\prime
}}-\varepsilon _{k^{\prime }}\right) }{\eta _{k_{1}k^{\prime }}\left( \eta
_{k_{1}k^{\prime }}+U\right) }-\frac{U\xi _{k^{\prime }-k+k_{1}}\delta
\left( \omega +\eta _{k_{1}k^{\prime }}-\varepsilon _{k^{\prime }}-U\right) 
}{\eta _{k_{1}k^{\prime }}\left( \eta _{k_{1}k^{\prime }}-U\right) }  \notag
\\
&&-\frac{U\xi _{-k}\delta \left( \omega +\eta _{k_{1}k^{\prime
}}-\varepsilon _{k^{\prime }}-\left( 1-n_{\sigma }\right) U\right) }{\left(
\eta _{k_{1}k^{\prime }}-\left( 1-n_{\sigma }\right) U\right) \left( \eta
_{k_{1}k^{\prime }}+n_{\sigma }U\right) }  \label{31a} \\
&&+\frac{U\xi _{k^{\prime }-k+k_{1}}\delta \left( \omega +\eta
_{k_{1}k^{\prime }}-\varepsilon _{k^{\prime }}-n_{\sigma }U\right) }{\left(
\eta _{k_{1}k^{\prime }}-n_{\sigma }U\right) \left( \eta _{k_{1}k^{\prime
}}+\left( 1-n_{\sigma }\right) U\right) }  \notag
\end{eqnarray}%
And the spectral density function $A_{\sigma }^{pair}\left( \omega ,k\right) 
$ is contributed by the electronic pairing parameters $\Delta _{k_{1}}$,%
\begin{equation}
A_{\sigma }^{pair}\left( \omega ,k\right) =\frac{2\pi }{\left( 1-n_{\sigma
}\right) N^{2}}\sum_{k^{\prime }k_{1}}\Lambda _{k_{1}k^{\prime }}\left(
\omega \right) \Delta _{k_{1}}^{\ast }\Delta _{k^{\prime }}  \label{32}
\end{equation}%
\begin{eqnarray}
\Lambda _{k_{1}k^{\prime }}\left( \omega \right) &=&A_{k_{1}k^{\prime
}}^{NL}\left( -k-k_{1}\right) \delta \left( \omega -\varepsilon
_{-k_{1}}\right) -B_{k_{1}k^{\prime }}^{NL}\left( -k-k_{1}\right) \delta
\left( \omega -\varepsilon _{-k_{1}}-U\right)  \notag \\
&&+\frac{U\xi _{-k-k_{1}-k^{\prime }}\delta \left( \omega -\varepsilon
_{k^{\prime }}\right) }{\eta _{k_{1}k^{\prime }}\left( \eta _{k_{1}k^{\prime
}}+U\right) }+\frac{U\xi _{-k}\delta \left( \omega -\varepsilon _{k^{\prime
}}-U\right) }{\eta _{k_{1}k^{\prime }}\left( \eta _{k_{1}k^{\prime
}}-U\right) }  \notag \\
&&-\frac{U\xi _{-k-k_{1}-k^{\prime }}\delta \left( \omega -\varepsilon
_{k^{\prime }}-\left( 1-n_{\sigma }\right) U\right) }{\left( \eta
_{k_{1}k^{\prime }}+n_{\sigma }U\right) \left( \eta _{k_{1}k^{\prime
}}-\left( 1-n_{\sigma }\right) U\right) }  \label{32a} \\
&&+\frac{U\xi _{-k}\delta \left( \omega -\varepsilon _{k^{\prime
}}-n_{\sigma }U\right) }{\left( \eta _{k_{1}k^{\prime }}-n_{\sigma }U\right)
\left( \eta _{k_{1}k^{\prime }}+\left( 1-n_{\sigma }\right) U\right) } 
\notag
\end{eqnarray}%
At the half filling, the spectral density function $A_{\sigma }^{Con}\left(
\omega ,k\right) $ does not be zero, while the spectral density function $%
A_{\sigma }^{Em}\left( \omega ,k\right) $ is zero. This means that at the
half filling there only emerge the collective modes corresponding to the
electronic spin density wave, and the uniform electronic pairing parameter $%
\Delta _{k\prime }$ is expectedly zero in the large $U$ limit.

At the half filling, the last two terms residing around in the middle of the
gap in the expressions of the functions $\Gamma _{k_{1}k^{\prime }}\left(
\omega \right) $ and $\Lambda _{k_{1}k^{\prime }}\left( \omega \right) $ are
small quantities because they are inverse proportion to $U$ in the large $U$
limit, and the contribution of the collective modes to the spectral density
function is not zero mainly around the lower and upper Hubbard bands. Of
course, for any electron density, the total spectral density function $%
A_{\sigma }\left( \omega ,k\right) $ must be positive, meanwhile it also
satisfy the sum rule in Eq.(\ref{26}).

The above properties of the spectral density function $A_{\sigma }\left(
\omega ,k\right) $ is based on the cut-off approximations in Eq.(\ref{13c}).
However, to high order approximations, if we solve the EOMs of the
correlation functions $F_{nliljq\sigma }^{(n_{\sigma }n_{\sigma }X_{%
\overline{\sigma }}^{\mp })}(t_{1},t_{2})$, under taking the cut-off
approximations similar to that in the Eq.(\ref{13c}) for the correlation
functions belonging to the level $L=4$ that emerge in the EOMs of the
correlation functions $F_{nliljq\sigma }^{(n_{\sigma }n_{\sigma }X_{%
\overline{\sigma }}^{\mp })}(t_{1},t_{2})$, we will find that the basic
behavior of the spectral density function $A_{\sigma }\left( \omega
,k\right) $ is unchanged, in which the total spectral weight of the emerged
states of electrons in the gap is also proportional to the hole doping
concentration $\delta $ in the large $U$ limit, that is independent of the
cut-off approximations in Eq.(\ref{13c}).

\section{Conclusions}

By introducing multipe-site correlation functions, we have proposed a
hierarchical Green function approach, and applied it to study a 2D square
lattice Hubbard model by solving the EOMs of one-particle Green function and
related multipe-site correlation functions. Under a cut-off approximation
for the correlation functions belonging to the level $L=3$ that emerge in
the EOMs of the multipe-site correlation functions $F_{niljq\sigma
}^{(n_{\sigma }X_{\overline{\sigma }}^{\mp })}(t_{1},t_{2})$. Then by using
the Fourier representation of these correlation functions, we have solved
this set of closed EOMs, and obtained an analytical expression of
one-particle Green function with possible order parameters emerging in the
ground state. With this one-particle Green function, we have calculated the
spectral density function of electrons, and found that besides usual two
main peaks corresponding to the lower and upper Hubbard bands in the
spectral density function, there emerge some novel states between these two
main peaks, and the total spectral weight of these emerged states is
proportional to the hole doping concentration $\delta $ for large $U$, that
is distinct from that one in a metal. Meanwhile, there also emerge some
collective modes related to possible charge/spin density wave and/or
electronic pairing density wave ordering states. However, at the half
filling, it is a Mott insulator with a gap that is in proportion to $U$ in
the large $U$ limit. The present results are completely consistent with the
spectroscopy observations of the cuprate superconductors in normal states%
\cite{8ad1,8ad2,8ad3}. On the other hand, the appearence of the static
correlation functions in the one-particle Green function can be used to
describe the intertwined orders observed in the normal state of the cuprate
superconductors. Moreover, the present approach can also be used to study
other quantum many particle systems, such as Anderson impurity model and
Heisenberg model.

\section{\protect\bigskip Acknowledgments}

This work is partially supported by the Fundamental Research Funds for the
Central Universities, and the Research Funds of Renmin University of China
(14XNLQ03), and by the National Basic Research Program of China,
No.2012CB921704.

\end{document}